\documentclass[onecolumn, draftclsnofoot, 12pt]{IEEEtran}
\usepackage{amsmath}
\usepackage{cite}
\usepackage{amssymb}
\usepackage{amsfonts}
\usepackage{algorithm}
\usepackage{dsfont}
\usepackage{graphicx}
\usepackage{epsfig}
\usepackage{subfigure}
\usepackage{psfrag}
\usepackage{xcolor}

\newtheorem{theorem}{\underline{Theorem}}
\newtheorem{proposition}{\underline{Proposition}}

\newtheorem{remark}{\underline{Remark}}
\newcommand{\mv}[1]{\mbox{\boldmath{$ #1 $}}}

\title{Multi-Beam UAV Communication in Cellular Uplink: Cooperative Interference Cancellation and Sum-Rate Maximization}
\author{Liang Liu, Shuowen Zhang, and Rui Zhang
\thanks{The authors are with the Department of Electrical and
Computer Engineering, National University of Singapore, Singapore
(e-mails: \{eleliu,elezhsh,elezhang\}@nus.edu.sg). Part of this paper was submitted to IEEE Global Communications Conference (Globecom), workshop on Wireless Networking and Control for Unmanned Autonomous Vehicles, 2018. }}

\begin{document}
\maketitle \thispagestyle{empty} \vspace{-0.3in}

\begin{abstract}
Integrating unmanned aerial vehicles (UAVs) into the cellular network as new aerial users is a promising solution to meet their ever-increasing communication demands in a plethora of applications. Due to the high UAV altitude, the channels between UAVs and the ground base stations (GBSs) are dominated by the strong line-of-sight (LoS) links, which brings both opportunities and challenges. On one hand, a UAV can communicate with a large number of GBSs at the same time, leading to a higher macro-diversity gain as compared to terrestrial users. However, on the other hand, severe interference may be generated to/from the GBSs in the uplink/downlink, which renders the interference management with coexisting terrestrial and aerial users a more challenging problem to solve. To deal with the above new trade-off, this paper studies the uplink communication from a multi-antenna UAV to a set of GBSs in its signal coverage region. Among these GBSs, we denote \emph{available GBSs} as the ones that do not serve any terrestrial users at the assigned resource block (RB) of the UAV, and \emph{occupied GBSs} as the rest that are serving their respectively associated terrestrial users in the same RB. We propose a new \emph{cooperative interference cancellation} strategy for the multi-beam UAV uplink communication, which aims to eliminate the co-channel interference at each of the occupied GBSs and in the meanwhile maximize the sum-rate to the available GBSs. Specifically, the multi-antenna UAV sends multiple data streams to selected available GBSs, which in turn forward their decoded data streams to their backhaul-connected occupied GBSs for interference cancellation. To draw useful insights and facilitate our proposed design, the maximum degrees-of-freedom (DoF) achievable by the multi-beam UAV communication for sum-rate maximization in the high signal-to-noise ratio (SNR) regime is first characterized, subject to the stringent constraint that all the occupied GBSs do not suffer from any interference in the UAV's uplink transmission. Then, based on the DoF-optimal  design, the achievable sum-rate at finite SNR is maximized, subject to given maximum allowable interference power constraints at each of the occupied GBSs. Numerical examples validate the DoF and sum-rate performance of our proposed designs, as compared to benchmark schemes with fully cooperative, local or no interference cancellation at the GBSs.
\end{abstract}


\section{Introduction}\label{sec:Introduction}
Recently, unmanned aerial vehicles (UAVs) or drones have found a wide range of applications in package delivery, video surveillance, remote sensing, aerial communication platform, and many others, thanks to their flexible deployment and high mobility \cite{Zhang18}. To embrace the upcoming era of ``internet-of-drones'', it is imperative to enable high-performance communications between UAVs and their ground users/pilots. Specifically, UAVs need to receive real-time control and command signals from the ground with high reliability to guarantee their operation safety, as well as to deliver mission-related payload data (e.g., high-resolution images/videos) to the ground with high rate. However, most UAVs in the current market communicate with the ground via simple point-to-point links over the unlicensed spectrum, which can only operate in the visual line-of-sight (VLoS) range and result in limited performance and security. To overcome this issue, a promising solution is \emph{cellular-enabled UAV communication}, which leverages the ground base stations (GBSs) in cellular networks for realizing high-performance UAV-ground communications \cite{LTE,Lin17,CellularConnected}. Compared to the existing approach, cellular-enabled UAV communication enables beyond visual line-of-sight (BVLoS) UAV communications with the ground users, and is envisioned to yield tremendous performance enhancement with today's mature cellular technology.
\begin{figure}[t]
  \centering
  \includegraphics[width=8cm]{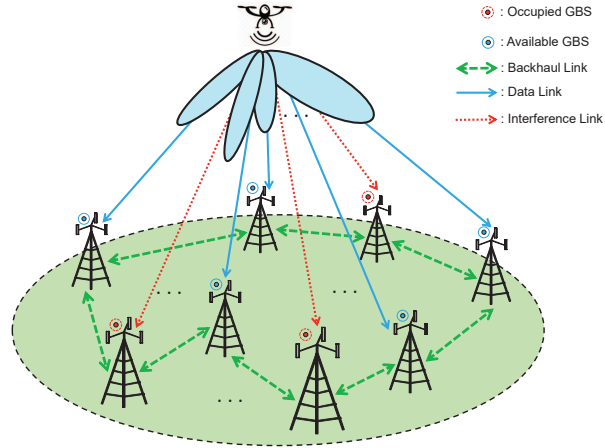}
  \caption{Illustration of the proposed multi-beam UAV communication in cellular uplink.}\label{system}
\end{figure}

Compared to traditional terrestrial users, UAVs possess dramatically distinct wireless channel characteristics with the GBSs. Different from the terrestrial channels that typically experience more significant attenuation over distance, shadowing and multi-path fading, the UAV-GBS channel is generally dominated by the strong \emph{line-of-sight (LoS)} link due to the high altitude of UAVs \cite{Qualcomm}. Hence, UAVs in general have better channel conditions with much more GBSs as compared to terrestrial users, which brings both opportunities and challenges to cellular communications. On one hand, the more available \emph{macro-diversity} in association with GBSs can be exploited to enhance the UAV communication performance in terms of reliability and throughput by connecting to the GBS with the best channel at each time instant \cite{Zhang18} or multiple GBSs at the same time, especially when the UAVs have multiple antennas. On the other hand, for the aerial and terrestrial users that simultaneously use the same time-frequency resource block (RB), severe \emph{co-channel interference} also occurs which may degrade significantly the communication performance of UAVs in the downlink as well as terrestrial users in the uplink \cite{Azari17,Nguyen18}.

To resolve the above fundamental trade-off in cellular-enabled UAV communication, we consider in this paper a new multi-beam UAV communication scenario in the cellular uplink, as shown in Fig. \ref{system}, where a UAV equipped with multiple antennas sends multiple data streams to its associated GBSs over a given RB assigned. By exploiting the macro-diversity, the multi-beam communication can help significantly improve the spectral efficiency for the UAVs to meet their high-rate payload data transmission requirement in the uplink. We divide the GBSs in the UAV's signal coverage region into two groups: \emph{occupied} GBSs each already having an associated terrestrial user to communicate in the same RB as the UAV, and \emph{available} GBSs which do not serve any terrestrial users in the uplink over this RB. Then, we aim to design the transmit beamforming at the UAV to deliver its messages to a selected set of available GBSs to maximize its uplink sum-rate, while in the meanwhile suppressing the interference caused to the occupied GBSs for protecting their terrestrial user communications.

\subsection{Prior Work}
Recently, the UAV uplink communication has appealed a lot of attention in the literature. In \cite{Larsson18,Larsson}, massive MIMO (multiple-input multiple-output) technique is employed at the GBS side to mitigate the interference between multiple UAVs in the uplink communications. However, interference control to the terrestrial communications is not considered in these works. On the other hand, some initial field tests have pointed out the severe interference generated to the terrestrial communications from the aerial users in the cellular uplink \cite{Lin17,Lin18}. However, these works do not propose new solutions to deal with this interference issue.

Despite the lack of studies on our considered UAV communication system in the cellular uplink, effective interference control strategies have been widely studied in the literature for terrestrial communications, including \emph{cognitive beamforming} \cite{Interference_Temperature,rui08}, \emph{non-orthogonal multiple access (NOMA)} \cite{NOMA,Ding}, \emph{coordinated multipoint (CoMP)} \cite{CoMP}, etc. An important but unaddressed question is whether these schemes can be applied to our considered system with both aerial and terrestrial users.

First, our considered system shares similarity with the cognitive radio (CR) network, in which a secondary user utilizes the transmit spectrum originally allocated to the existing primary users for communications \cite{Interference_Temperature,rui08}. In these works, beamforming design at the secondary user is optimized to maximize its transmit rate under the constraint that the interference seen by each primary receiver is no larger than a threshold known as ``interference temperature''. However, the above cognitive beamforming strategy may be ineffective in our considered multi-beam UAV communication in cellular uplink. This is because due to the large frequency reuse factor in today's cellular networks as well as the LoS-dominant UAV-ground links, the number of occupied GBSs may practically far exceed the number of antennas at the UAV, which yields limited or even zero degrees-of-freedom (DoF) for the UAV's multi-beam uplink communication after projecting the beamspace to the orthogonal space of all occupied GBSs' channels. As a result, the spatial multiplexing gain of the multi-beam transmission cannot be realized, {\hbox{even when the number of available GBSs is large.}}

Another effective approach for interference control is NOMA \cite{NOMA,Ding}. Under this strategy, each receiver can employ the successive interference cancellation technique to first decode the strong interference and then subtract the interference for useful message decoding. In our considered system, NOMA-based beamforming allows the UAV to either generate strong interference to occupied GBSs such that they can decode and cancel it, or suppress the interference to protect the occupied GBSs. Such a flexibility promises a reasonable sum-rate gain of the NOMA strategy over the cognitive beamforming at the medium signal-to-noise ratio (SNR) regime. However, this gain is expected to be marginal at the high SNR regime since successive interference cancellation cannot improve the DoF of the system \cite{Cover}.

On the other hand, CoMP can significantly improve the DoF of the system by leveraging the cooperations between different GBSs. For example, in cloud radio access network (C-RAN) \cite{CRAN,Liang15}, a set of GBSs are connected to a central processor via the fronthaul links. By a joint decoding over the signals received at all the GBSs, the maximum DoF can be achieved assuming perfect fronthaul links. However, CoMP requires substantial overheads over the backhaul/fronthaul links as the received signals at all GBSs (both occupied and available in our considered system) need to be quantized and transmitted. Moreover, since multi-hop routing is in general necessary, CoMP inevitably introduces a long delay for information decoding, making it not suitable for delay-sensitive UAV applications.

To summarize, due to the unique feature of the multi-beam UAV communication in cellular uplink, i.e., a large number of (available and occupied) GBSs are in the UAV's signal coverage region, the existing interference management strategies for terrestrial communications cannot be directly used in our considered system, and innovative solutions are needed to tackle this challenge.

\subsection{Main Contributions}
The contributions of this paper are summarized as follows.

First, to harvest the gain of cooperative decoding but with low implementation cost and low delay, we propose a novel \emph{cooperative interference cancellation} strategy, by exploiting the existing backhaul connections between adjacent GBSs, e.g., Xhaul \cite{Xhaul}, in cellular networks. Specifically, we first select a subset of the available GBSs, each for decoding one data stream sent from the UAV; then, we let these GBSs forward the decoded messages to their backhaul-connected occupied GBSs for the latter to cancel the UAV's interference. In this regard, for each data stream sent by the UAV, the number of occupied GBSs that still require interference nulling via cognitive beamforming at the UAV is significantly reduced. As a result, more data streams can be transmitted under the interference temperature constraints of the occupied GBSs as compared to the cognitive beamforming and NOMA strategies. Moreover, this scheme merely requires backhaul connections between adjacent GBSs, and is more practically appealing as compared to the CoMP strategy.

Next, due to the typical high-SNR communication links between the UAV and GBSs resulting from the strong LoS-dominant channels as well as for drawing useful insights, we first provide a DoF analysis for the considered multi-beam UAV uplink communication system with GBSs' cooperative interference cancellation. Specifically, we derive the maximum achievable DoF with our proposed strategy for UAV's sum-rate maximization subject to the stringent constraint on no harmful interference to existing terrestrial uplink communications, by jointly optimizing the number of independent data streams sent by the UAV as well as their associated available GBSs. We also present a zero-forcing (ZF) based transmit beamforming design to achieve the maximum DoF. It is shown analytically that our strategy can achieve higher DoF compared to the cognitive beamforming strategy as well as the NOMA strategy.

Furthermore, this paper also considers the UAV sum-rate maximization problem at the finite SNR regime, under the interference temperature constraints for protecting the occupied GBSs. Utilizing the DoF-optimal data stream association solution, we reveal that our considered system reduces to a multi-group multicast channel \cite{multicasting}, and thus the existing beamforming design for sum-rate maximization in the broadcast channel \cite{WMMSE,optimality} cannot be applied. Nevertheless, this paper proposes an efficient beamforming design algorithm based on the successive convex approximation technique for solving the formulated problem locally optimally. Numerical results are provided to show the significant sum-rate gain over the cognitive beamforming strategy.

\subsection{Organization}
The rest of this paper is organized as follows.
Section \ref{sec:System Model} describes the system model for our considered multi-beam UAV communication in cellular uplink.
Section \ref{sec:Cooperative Interference Cancellation} introduces the proposed cooperative interference cancellation strategy to protect the occupied GBSs.
Section \ref{sec:UAV Transmission Rate Maximization with Interference Control} formulates the sum-rate maximization problem under the interference temperature constraint.
Section \ref{sec:Proposed Solution} characterizes the DoF of the considered system by solving the formulated problem in the infinite SNR regime.
Section \ref{sec:Proposed Solution in Finite SNR Regime} proposes an efficient algorithm to solve the formulated problem in the finite SNR regime. Section \ref{sec:Numerical Examples} provides the numerical simulation results to evaluate the performance of the proposed cooperative interference cancellation strategy. Finally, Section \ref{sec:Conclusion} concludes the paper and outlines the future research directions.

{\it Notation}: Scalars are denoted by lower-case letters, vectors
by bold-face lower-case letters, and matrices by
bold-face upper-case letters. $\mv{I}$ and $\mv{0}$  denote an
identity matrix and an all-zero matrix, respectively, with
appropriate dimensions. For a matrix
$\mv{M}$ of arbitrary size, $\mv{M}^{H}$ denotes its
conjugate transpose. The
distribution of a circularly symmetric complex Gaussian (CSCG) random vector with mean $\mv{x}$ and
covariance matrix $\mv{\Sigma}$ is denoted by
$\mathcal{CN}(\mv{x},\mv{\Sigma})$; and $\sim$ stands for
``distributed as''. $\mathbb{C}^{x \times y}$ denotes the space of
$x\times y$ complex matrices. $\|\mv{x}\|$ denotes the Euclidean norm of a complex vector
$\mv{x}$.

\section{System Model}\label{sec:System Model}

As shown in Fig. \ref{system}, we consider the uplink communication in a cellular network consisting of one UAV equipped with $M\geq1$ antennas and $N$ GBSs in the UAV's signal coverage region denoted by the set $\mathcal{N}=\{1,\cdots,N\}$. We assume that each GBS has a fixed beam pattern for the aerial user, thus can be equivalently viewed as being equipped with one single antenna for the purpose of exposition.\footnote{In practice, the fixed beam pattern may be resulted from the fact that in the current 4G-LTE (long-term evolution) cellular network, GBS antennas are generally tilted downwards for mitigating inter-cell interference of terrestrial communications, thus aerial users can only be served by the sidelobes.} The general case of multi-antenna GBSs with flexible three-dimensional (3D) beamforming will be considered in future work.

Denote $\mv{h}_n\in \mathbb{C}^{M\times 1}$ as the effective uplink channel vector from the UAV to GBS $n$, $n=1,\cdots,N$; and $h_{n,m}$ as the channel coefficient from the $m$th antenna of the UAV to GBS $n$ such that $\mv{h}_n=[h_{n,1},\cdots,h_{n,M}]^H$. In this paper, we consider that the UAV is at a fixed location (e.g., when hovering), where $d_n$ in meters (m) denotes the distance from the UAV to GBS $n$. Moreover, we consider the Rician fading channel model, and the channel from the UAV to GBS $n$ is given by
\begin{align}\label{eqn:Rician channel}
\mv{h}_n=\sqrt{\frac{\tau_0}{d_n^2}}\left(\sqrt{\frac{\lambda}{\lambda+1}}\hat{\mv{h}}_n+\sqrt{\frac{1}{\lambda+1}}\tilde{\mv{h}}_n\right),
\end{align}where $\tau_0$ denotes the channel power gain at the reference distance $d_0=1$ m; $\hat{\mv{h}}_n\in \mathbb{C}^{M\times 1}$ with $\|\hat{\mv{h}}\|=1$ denotes the LoS channel component; $\tilde{\mv{h}}_n\in \mathbb{C}^{M\times 1}$ with $\tilde{\mv{h}}\sim \mathcal{CN}(\mv{0},\mv{I})$ denotes the Rayleigh fading channel component; and $\lambda\geq 0$ is the Rician factor specifying the power ratio between the LoS and Rayleigh fading components in $\mv{h}_n$. Note that under the above Rician fading channel model, the channel vectors between the multi-antenna UAV and any $N',\ N'\leq \min(M,N)$, GBSs are linearly independent with probability one. We further assume that all ${\mv{h}}_n$'s are known at the UAV.\footnote{In practice, the UAV may send $\min(M,N)$ orthogonal pilots for channel training, and the GBSs can feed back their channels to the UAV.}

The UAV employs the multi-beam scheme to transmit its messages to some selected (available) GBSs at an assigned RB, which is assumed given in this paper; while in practice it can be dynamically assigned in the network. The transmit signal of the UAV is expressed as
\begin{equation}\label{eqn:transmit signal}
\mv{x}=\sum\limits_{j=1}^J\mv{w}_js_j,
\end{equation}
where $J$ denotes the number of independent data streams sent by the UAV, $s_j\sim \mathcal{CN}(0,1)$ denotes the message for the $j$th data stream, and $\mv{w}_j\in \mathbb{C}^{M\times 1}$ denotes the transmit beamforming vector for $s_j$. The received signal at the $n$th GBS at this given RB is thus given by
\begin{equation}\label{eqn:received signal GBS}
y_{n}=\mv{h}_{n}^H\sum\limits_{j=1}^J\mv{w}_js_j+S_n+z_{n}, ~~~ n\in \mathcal{N},
\end{equation}
where $S_n$ denotes the received signal from the terrestrial user that is transmitting at the same RB as the UAV in cell $n$, and $z_n\sim \mathcal{CN}(0,\sigma_n^2)$ denotes the aggregated noise due to the interference generated by the terrestrial users that are transmitting at the same RB as the UAV in the other cells as well as the additive white Gaussian noise (AWGN) at GBS $n$. It is worth noting that $S_n\neq 0$ if the UAV's RB is used by GBS $n$ to serve a terrestrial user in its cell, and $S_n=0$ otherwise.

\begin{figure}
\begin{center}
\includegraphics[width=8cm]{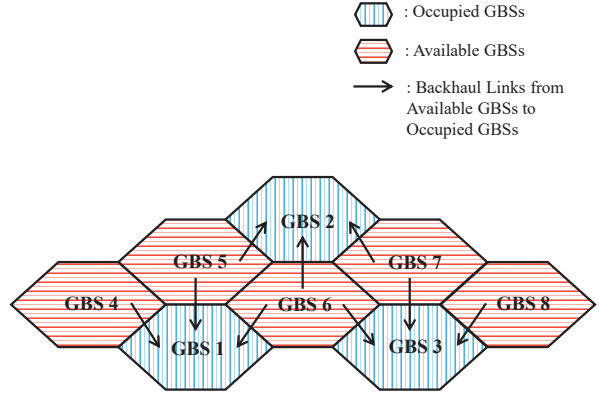}
\end{center}
\vspace{-5mm}
\caption{An example of the cellular network topology with fixed frequency reuse.}\label{cell}
\vspace{-7mm}
\end{figure}

Suppose that among the $N$ GBSs, $N_1<N$ GBSs are currently serving their terrestrial users at the same RB as the UAV (denoted as \emph{occupied} GBSs), while $N_2=N-N_1$ GBSs are not using this RB to serve any terrestrial users (denoted as \emph{available} GBSs). For convenience, we define $\mathcal{N}_1=\{1,\cdots,N_1\}$ as the set of $N_1$ occupied GBSs, and $\mathcal{N}_2=\{N_1+1,\cdots,N\}$ as the set of $N_2$ available GBSs. In traditional cellular networks with a fixed frequency reuse factor, the occupied GBSs may correspond to those which are assigned the same frequency band for serving some terrestrial users in their associated cells, which contains the sub-band of the RB assigned to the UAV. Then  the available GBSs are those assigned with other orthogonal frequency bands and thus can serve the UAV opportunistically without affecting the uplink communications of their own terrestrial users (at different frequency from that of the UAV). As a result, the cellular spectrum can be efficiently utilized to serve the UAV users by exploiting the macro-diversity, provided that the strong interference to the occupied GBSs is effectively mitigated. Notice that this leads to an interesting new cell association strategy for the UAV's uplink communication, where its associated (available) GBSs are simultaneously serving their terrestrial users in other frequency bands, which is in sharp contrast to the conventional cell association of terrestrial users. For example, under the cellular network topology shown in Fig. \ref{cell}, GBSs in the set $\mathcal{N}_1=\{1,2,3\}$ are occupied GBSs assigned with frequency band $B_1$, and those in the set $\mathcal{N}_2=\{4,5,6,7,8\}$ are available GBSs assigned with orthogonal bands $B_2$, $B_3$, etc.

Since the available GBSs are not using the same RB as the UAV, only the terrestrial users served by the occupied GBSs may generate inter-cell interference to the available GBSs at this particular RB (modeled in $z_n$ for $n\in \mathcal{N}_2$ given in (\ref{eqn:received signal GBS})). In this paper, we assume that such terrestrial interference has been well mitigated due to the non-LoS terrestrial channel fading and the inter-cell interference coordination (ICIC) techniques (such as cooperative RB allocation, beamforming, power control, etc.) employed by the GBSs, and thus any leakage interference is assumed to be much weaker and almost negligible as compared to the UAV's uplink signal received at each available GBSs. As a result, the UAV can opportunistically send its data streams to any available GBSs with negligible interference from terrestrial communications. However, the interference caused by the UAV to the terrestrial uplink communications at the occupied GBSs (as given in (\ref{eqn:received signal GBS}) for $n\in N_1$) is strong and cannot be ignored.

Since each GBS equivalently has one antenna due to fixed beam pattern, the UAV can send at most one data stream to each available GBS. Define $\Lambda_j\subseteq \mathcal{N}_2$ as the set of available GBSs that decode $s_j$, $j=1,\cdots,J$, where $\Lambda_j\neq \emptyset$, $\forall j$, and $\Lambda_j\bigcap \Lambda_i=\emptyset$, $\forall i\neq j$. Then for an available GBS $n_2\in \Lambda_j$, the signal-to-interference-plus-noise ratio (SINR) for decoding the $j$th data stream of the UAV is
\begin{align}\label{eqn:SINR}
\gamma_{n_2,j}=\frac{|\mv{h}_{n_2}^H\mv{w}_j|^2}{\sum\limits_{i\neq j}|\mv{h}_{n_2}^H\mv{w}_i|^2+\sigma_{n_2}^2}, ~~~ n_2\in \Lambda_j.
\end{align}Accordingly, the achievable rate in bits/second/Herz (bps/Hz) for multicasting data stream $j$ to its associated available GBSs is given by
\begin{align}\label{eqn:rate}
R_j=\min\limits_{n_2\in \Lambda_j} \log_2\left(1+\frac{|\mv{h}_{n_2}^H\mv{w}_j|^2}{\sum\limits_{i\neq j}|\mv{h}_{n_2}^H\mv{w}_i|^2+\sigma_{n_2}^2}\right), ~~~ \forall j.
\end{align}

On the other hand, as shown in (\ref{eqn:received signal GBS}), each occupied GBS $n_1$ needs to decode the message $S_{n_1}$ sent by its served terrestrial user. Due to the strong LoS channels between the UAV and GBSs, the UAV should carefully control its interference to each occupied GBS to protect its terrestrial uplink communication. In this paper, we consider that the interference power generated at each occupied GBS $n_1$ should be no larger than a given interference temperature constraint \cite{Interference_Temperature}, which is denoted by $\Theta_{n_1}\geq 0$. Note that if the UAV simply leverages the transmit beamforming technique to control the interference at all the occupied GBSs, according to (\ref{eqn:received signal GBS}), the following conditions need to be satisfied:
\begin{equation}\label{eqn:interference occupied}
\sum\limits_{j=1}^J|\mv{h}_{n_1}^H\mv{w}_j|^2\leq \Theta_{n_1}, ~~~ \forall n_1\in \mathcal{N}_1.
\end{equation}
However, in our considered system, there may not even exist a feasible beamforming solution to satisfy (\ref{eqn:interference occupied}) when $\|\mv{w}_j\|>0$, $\forall j$, and $\Theta_{n_1}$'s are sufficiently small. For example, the number of occupied GBSs in practice can be larger than the number of antennas at the UAV due to the LoS channels, i.e., $N_1\geq M$. In the extreme case when the interference from the UAV to the occupied GBSs needs to be totally mitigated, i.e., $\Theta_{n_1}=0$, $\forall n_1\in \mathcal{N}_1$, we must have $|\mv{h}_{n_1}^H\mv{w}_j|^2=0$, $\forall n_1$, $\forall j$. With $N_1\geq M$, almost surely there does not exist a feasible non-zero beamforming solution for the UAV to satisfy the above equations. Therefore, we are motivated to propose a new strategy in the next section for more effective air-to-ground interference control such that the interference temperature constraints can be more easily satisfied at the occupied GBSs.

\section{Cooperative Interference Cancellation}\label{sec:Cooperative Interference Cancellation}

In this section, we propose a novel cooperative interference cancellation strategy, which can effectively mitigate the interference caused by the uplink multi-beam UAV communication to the uplink terrestrial communications at the occupied GBSs, even in the case when the number of occupied GBSs is large and the interference threshold at each occupied GBS is small. Note that the proposed strategy is motivated by the recent developments of the Xhaul structure \cite{Xhaul} in cellular networks, where adjacent GBSs are connected by the backhaul links such that they can exchange information efficiently.

Specifically, since each of the available GBSs can opportunistically decode one data stream from the UAV, they can forward their decoded data streams to their connected occupied GBSs over the backhaul links such that these occupied GBSs can cancel the interference caused by the UAV's data streams. For any occupied GBS $n_1\in \mathcal{N}_1$, define $\Phi_{n_1}\subseteq \mathcal{N}_2$ as the set of available GBSs that have {\it{one-hop}} backhaul connections with the occupied GBS $n_1$.\footnote{In this paper, to minimize the processing delay to cater for the high UAV mobility in practice, we assume that each available GBS can merely forward its decoded message to an occupied GBS that is backhaul-connected to it without any intermediate GBSs (hops).} Moreover, define
\begin{equation}\label{eqn:set of available GBSs with data stream j}
\Omega_{n_1,j}=\Phi_{n_1}\bigcap \Lambda_j, ~~~ n_1\in \mathcal{N}_1, ~ j=1,\cdots,J,
\end{equation}
as the set of available GBSs that are connected to the occupied GBS $n_1$ and can decode the message from the $j$th data stream, i.e., $s_j$. If $\Omega_{n_1,j}\neq \emptyset$, then the available GBSs in $\Omega_{n_1,j}$ can send the decoded $s_j$ to the occupied GBS $n_1$ via the backhaul links, which then subtracts the interference caused by $s_j$ from its received signal (assuming that the corresponding effective channel coefficient ${\mv{h}}_{n_1}^H{\mv{w}}_j$ is estimated at GBS $n_1$); otherwise, if $\Omega_{n_1,j}=\emptyset$, then no available GBS can help the occupied GBS $n_1$ to cancel the interference caused by $s_j$. Define
\begin{equation}\label{eqn:interference data}
\Gamma_{n_1}=\{j:\Omega_{n_1,j}=\emptyset, j=1,\cdots,J\}, ~~~ n_1\in \mathcal{N}_1,
\end{equation}
as the set of data streams that the occupied GBS $n_1$ cannot obtain from its connected available GBSs. Then for the occupied GBSs, the received signal model given in (\ref{eqn:received signal GBS}) reduces to the following form after cooperative interference cancellation:
\begin{align}\label{eqn:received signal GBS interference cancellation}
y_{n_1}=\mv{h}_{n_1}^H\sum\limits_{j\in\Gamma_{n_1}} \mv{w}_js_j+S_n+z_{n_1}, ~~~ n_1\in \mathcal{N}_1.
\end{align}Therefore, the UAV merely needs to utilize the beamforming technique to control the residual interference caused by the data streams $s_j$'s, $\forall j\in \Gamma_{n_1}$, such that the following conditions hold:
\begin{equation}\label{eqn:remaining interference}
\sum\limits_{j\in \Gamma_{n_1}} |\mv{h}_{n_1}^H\mv{w}_j|^2\leq \Theta_{n_1}, ~~~ \forall n_1\in \mathcal{N}_1.
\end{equation}
The above strategy is referred to as \emph{cooperative interference cancellation}, since the available GBSs are utilized to forward some decoded messages to their connected occupied GBSs for interference cancellation. Note that to fully utilize the backhaul connections for cooperative interference cancellation, the UAV usually needs to multicast each data stream $j$ to more than one available GBSs denoted by the set $\Lambda_j$ as shown in (\ref{eqn:received signal GBS})--(\ref{eqn:rate}), whereas in the case without cooperative interference cancellation for the occupied GBSs, it is sufficient to send each data stream to one available GBS only.

By comparing (\ref{eqn:interference occupied}) and (\ref{eqn:remaining interference}), it is observed that with the above strategy, the interference temperature constraint is generally easier to be satisfied since $\sum_{j\in \Gamma_{n_1}} |\mv{h}_{n_1}^H\mv{w}_j|^2\leq \sum_{j=1}^J|\mv{h}_{n_1}^H\mv{w}_j|^2$, $\forall n_1$. Take the extreme case with zero-interference, i.e., $\Theta_{n_1}=0$, $\forall n_1$, or equivalently, $|\mv{h}_{n_1}^H\mv{w}_j|^2=0$, $\forall n_1$, $\forall j$, as an example. Consider the network topology shown in Fig. \ref{cell}, where we have $N_1=3$, $\Phi_{1}=\{4,5,6\}$, $\Phi_2=\{5,6,7\}$, and $\Phi_3=\{6,7,8\}$. Suppose that the UAV has $M=2$ antennas and its channels to occupied GBSs 1, 2, and 3 are linearly independent with each other (which is true under our considered Rician fading channel model (\ref{eqn:Rician channel})). Without the use of the proposed cooperative interference cancellation strategy, no data stream can be transmitted to the available GBSs without generating any interference to all the three occupied GBSs, since in this case the null space of the space spanned by $\mv{h}_1$, $\mv{h}_2$, and $\mv{h}_3$, which are linearly independent to each other, does not exist. However, with the proposed strategy, at least one data stream can be sent to GBS 6, which then forwards this data stream to occupied GBSs 1, 2, and 3 such that interference cancellation can be performed even without the need of UAV's ZF beamforming.

\section{Sum-Rate Maximization under Interference Constraints}\label{sec:UAV Transmission Rate Maximization with Interference Control}
Under the cooperative interference cancellation strategy proposed in Section \ref{sec:Cooperative Interference Cancellation}, in this paper we aim to maximize the UAV's sum-rate to the available GBSs in the uplink, i.e., $\sum_j R_j$ with $R_j$ given in (\ref{eqn:rate}), subject to the interference temperature constraints (\ref{eqn:remaining interference}). To achieve this goal, we need to jointly design the number of transmit data streams, i.e., $J$, the designated available GBSs of each data stream $j$, i.e., $\Lambda_j$, the corresponding beamforming vector and power allocation for multicasting each data stream $j$ to the available GBSs $\Lambda_j$, i.e., $\mv{w}_j$, as well as the multicasting rate for each data stream $j$, i.e., $R_j$. To summarize, the UAV's sum-rate maximization problem subject to the interference temperature constraints can be formulated as
\begin{subequations}\label{eqn:P1}\begin{align}
\mathop{ \underset{J,\{\Lambda_j,\mv{w}_j,R_j\}}{\max}} & ~ \sum\limits_{j=1}^J  R_j \label{eqn:problem}\\
\mathrm{s.t.} \ \ \ & \log_2(1+\gamma_{n_2,j})\geq R_j, ~~~ \forall n_2\in \Lambda_j, ~ j=1,\cdots,J,  \label{eqn:SINR constraint} \\ & \sum\limits_{j\in \Gamma_{n_1}}|\mv{h}_{n_1}^H\mv{w}_j|^2\leq \Theta_{n_1}, ~~~ \forall n_1\in \mathcal{N}_1, \label{eqn:interference temperature constraint} \\ & \Lambda_j\neq \emptyset, ~~~ j=1,\cdots,J, \label{eqn:message constraint} \\ & \Lambda_j\bigcap \Lambda_i =\emptyset, ~~~ i, j=1,\cdots,J ~ {\rm with} ~ i\neq j, \label{eqn:one data stream} \\ & \sum\limits_{j=1}^J\|\mv{w}_j\|^2 \leq P, \label{eqn:power constraint}
\end{align}\end{subequations}where (\ref{eqn:SINR constraint}) is due to the definition of multicasting rate given in (\ref{eqn:rate}), (\ref{eqn:interference temperature constraint}) is the interference temperature constraint to each occupied GBS after (possible) cooperative interference cancellation, (\ref{eqn:message constraint}) guarantees that each data stream needs to be sent to at least one available GBS, (\ref{eqn:one data stream}) guarantees that each available GBS can decode one data stream at most, and (\ref{eqn:power constraint}) guarantees that the total transmit power of the UAV is no larger than a power constraint denoted by $P$.

To find the optimal solution to the above problem, we need to first perform exhaustive search over $J$, with $1\leq J \leq \min(M,N)$; and for each given $J$, we need to conduct an exhaustive search over all the feasible data stream association with the available GBSs $\Lambda_j$'s; while given any assignment $\Lambda_j$'s, we need to optimize the beamforming vectors $\mv{w}_j$'s to maximize the sum-rate of all the UAV's data streams. At last, we select the optimal value of $J$ (and the corresponding optimal data stream association and optimal beamforming vectors) that results in the largest sum-rate. It is known that even with given $J$ and $\Lambda_j$'s, the sum-rate maximization problem via beamforming optimization is NP-hard \cite{beamforming}. As a result, if we jointly optimize $\Lambda_j$'s and $\mv{w}_j$'s directly for each given $J$, we need to solve an NP-hard beamforming problem for a very large number of times.

To draw useful insight to our formulated problem and facilitate low-complexity algorithm design, we first study the optimal solution to the above problem in the asymptotic regime when the transmit power of the UAV goes to infinity, i.e., $P\rightarrow \infty$. In this case, problem (\ref{eqn:P1}) reduces to the an equivalent DoF maximization problem with ZF constraints. Such an asymptotic analysis enables us to see more clearly the rate gain of the proposed cooperative interference cancellation strategy compared to the benchmark schemes with only local or no interference cancellation at the occupied GBSs, as will be shown in Section \ref{sec:Proposed Solution}. Moreover, as will be shown in Section \ref{sec:Proposed Solution in Finite SNR Regime}, based on the obtained DoF-optimal solutions for $\Lambda_j$'s, we propose an efficient beamforming design for problem (\ref{eqn:P1}) to maximize the sum-rate of the UAV in the finite SNR regime.

\begin{remark}\label{remark1}
It is worth noting that without the use of cooperative interference cancellation, we can only design the beamforming vectors of the UAV to control its interference to the occupied GBSs as shown in (\ref{eqn:interference occupied}) and maximize its transmission rate to the available GBSs. Such a beamforming design problem has been considered in the CR network where a secondary user aims to maximize its transmission rate over a shared channel with a set of primary users subject to given interference temperature constraints at their receivers \cite{Interference_Temperature}. For convenience, we refer to the above scheme as cognitive beamforming.

\end{remark}

\section{DoF Analysis}\label{sec:Proposed Solution}

Note that in the high-SNR regime, the DoF is a fundamental characterization of the achievable rate in MIMO communication systems \cite{Tse}. Due to the strong LoS channels between the UAV and GBSs, the high-SNR assumption is practically valid in our considered system. As a result, in this section we study the maximum DoF of the multi-beam UAV  uplink communication by our proposed strategy and compare it with that of other techniques such as CoMP/C-RAN, NOMA, and cognitive beamforming. The obtained design that maximizes the DoF will also be useful to our proposed solution for problem (\ref{eqn:P1}) in the finite SNR regime in Section \ref{sec:Proposed Solution in Finite SNR Regime}.

\subsection{Problem Formulation}\label{sec:Optimization 1}

The DoF represents the rate of growth for the network capacity with respect to the logarithm of the SNR \cite{Jafar08}. In our considered system, a DoF of $J$ is achievable if $J$ data streams can be transmitted from the UAV such that each data stream $j$ can be decoded by at least one available GBS, while its interference at the occupied GBSs as well as other available GBSs that decode the other data streams is zero.\footnote{Note that the interference temperature constraint at each occupied GBS is in general $\Theta_{n_1}\geq 0$ in problem (\ref{eqn:P1}). However, as $P$ goes to infinity, the interference of each data stream at each occupied GBS can be either zero or infinity. As a result, the only way to satisfy these interference temperature constraints is to ensure zero-interference at each of the occupied GBSs.} Under our proposed cooperative interference cancellation strategy, zero interference of $s_j$ at any occupied GBS can be achieved by either aligning $\mv{w}_j$ to the null space of this GBS's channel or selecting one of its connected available GBSs to decode $s_j$ for interference cancellation. For any data stream $j$, define
\begin{equation}\label{eqn:occupied GBS without data stream j}
\Psi_j\!=\!\{n_1\!:\!\Phi_{n_1}\bigcap \Lambda_j\!=\!\emptyset, n_1=1,\cdots\!,N_1\}, j=1,\cdots\!,J,
\end{equation}
as the set of occupied GBSs that cannot receive this data stream from its connected available GBSs. Then, as $P\rightarrow \infty$, problem (\ref{eqn:P1}) reduces to the following DoF maximization problem.
\begin{subequations}\label{eqn:P2}\begin{align}
\mathop{ \underset{J,\{\Lambda_j,\mv{w}_j\}}{\max}} & ~ J \label{eqn:problem 1}\\
\mathrm{s.t.} \ \ \ & \mv{h}_{n_2}^H\mv{w}_j\neq 0, ~~~ \forall n_2\in \Lambda_j, ~ j=1,\cdots,J, \label{eqn:DoF 1} \\ & \mv{h}_{n_2}^H\mv{w}_j=0, ~~~ \forall n_2\in \bigcup\limits_{i\neq j}\Lambda_i, ~ j=1,\cdots,J. \label{eqn:DoF 3} \\ & \mv{h}_{n_1}^H\mv{w}_j=0, ~~~ \forall n_1\in \Psi_j, ~ j=1,\cdots,J, \label{eqn:DoF 2} \\ & ({\rm \ref{eqn:message constraint}}), ~ ({\rm \ref{eqn:one data stream}}).
\end{align}\end{subequations}In the above problem, (\ref{eqn:DoF 1}) and (\ref{eqn:DoF 3}) guarantee that each $s_j$ can be decoded by the available GBSs in the set $\Lambda_j$ with an infinite SNR \cite{Jafar08}, and (\ref{eqn:DoF 2}) guarantees that the interference caused by each $j$th data stream at any occupied GBS that cannot receive it from the connected available GBSs (hence cannot cancel its interference) is zero.

The above DoF maximization problem can be solved as follows. First, given any $J$, we check whether there exists a feasible solution of $\Lambda_j$'s and $\mv{w}_j$'s to satisfy the conditions in problem (\ref{eqn:P2}). Then, the maximum achievable DoF of problem (\ref{eqn:P2}) can be obtained by a bisection search over the interval $[1,\min(M,N_2)]$, since in the most favorite case without any occupied GBSs, the maximum DoF is $\min(M,N_2)$.

\subsection{Feasibility Check}
Note that the key to solve problem (\ref{eqn:P2}) via the bisection method lies in how to check whether a DoF of $J$ is feasible or not given any network setup. The following theorem simplifies the feasibility check problem by characterizing the achievable DoF merely as a function of $\Lambda_j$'s.

\begin{theorem}\label{theorem1}
Under the Rician channel model (\ref{eqn:Rician channel}), a DoF of $J\leq \min(M,N_2)$ is achievable if and only if we can find a solution of $\Lambda_j$'s, $j=1,\cdots,J$, which satisfies the following conditions:
\begin{align}
& |\Psi_j|+\sum\limits_{i\neq j}|\Lambda_i|<M, ~~~ j=1,\cdots,J, \label{eqn:null space} \\
& ({\rm \ref{eqn:message constraint}}), ~ ({\rm \ref{eqn:one data stream}}). \nonumber
\end{align}
\end{theorem}

\begin{IEEEproof}
Please refer to Appendix \ref{appendix1}.
\end{IEEEproof}

The intuition of Theorem \ref{theorem1} is that there exists a feasible beamforming solution to satisfy (\ref{eqn:DoF 1}) and (\ref{eqn:DoF 3}) if and only if the data stream association $\Lambda_j$'s satisfy (\ref{eqn:null space}). As a result, the achievable DoF only depends on $\Lambda_j$'s. Based on the DoF characterization of the proposed strategy shown in Theorem \ref{theorem1}, we formulate the following feasibility problem to check whether a DoF $J$ is achievable by optimizing the association between data streams and available GBSs:

\begin{subequations}\label{eqn:P3}\begin{align}
\mathop{\mathrm{find}} & ~ \Lambda_1, ~ \cdots, ~ \Lambda_J  \\
\mathrm{s.t.} & ~ ({\rm \ref{eqn:message constraint}}), ~ ({\rm \ref{eqn:one data stream}}), ~ ({\rm \ref{eqn:null space}}).
\end{align}\end{subequations}

Compared to the original DoF feasibility check via problem (\ref{eqn:P2}), the number of variables in problem (\ref{eqn:P3}) is reduced since beamforming design is not involved. Moreover, the number of possible data stream association solutions of $\Lambda_j$'s is finite. As a result, we can solve problem (\ref{eqn:P3}) via an exhaustive search as follows. First, we list all the possible solutions of $\Lambda_j$'s that satisfy (\ref{eqn:one data stream}). Since each available GBS can be assigned to decode either one of the $J$ data streams or nothing, there are $(J+1)^{N_2}$ possible solutions of $\Lambda_j$'s in total. Then, we check whether there exists a solution among all that satisfies (\ref{eqn:message constraint}) and (\ref{eqn:null space}). If so, we say that a DoF of $J$ is achievable; otherwise, we claim that a DoF of $J$ is not achievable.

\subsection{DoF-Optimal Beamforming Design}\label{sec:Optimization 2}

In this subsection, we present a ZF-based transmit beamforming solution to problem (\ref{eqn:P2}) that can achieve the maximum DoF $J$ with the obtained data stream-GBS association $\Lambda_j$'s obtained in the preceding subsection. For ease of exposition, we assume that the set of available GBSs that receive the $j$th data stream is given by $\Lambda_j\!=\!\{N_1+\sum_{i=1}^{j-1}|\Lambda_i|+1,\!\cdots\!,N_1+\sum_{i=1}^{j}|\Lambda_i|\}$. Define ${\mv{H}}_{j}^{\mathrm{A}}\!=\!\left[{\mv{h}}_{N_1+\sum_{i=1}^{j-1}|\Lambda_i|+1},\!\cdots\!,{\mv{h}}_{N_1+\sum_{i=1}^{j}|\Lambda_i|}\right]\!\in\! \mathbb{C}^{M\times |\Lambda_j|}$ as the aggregate channel matrix from the UAV to all the available GBSs that receive the $j$th data stream, and ${\mv{H}}_{[-j]}^{\mathrm{A}}\!=\![{\mv{H}}_{1}^{\mathrm{A}},\cdots,{\mv{H}}_{j-1}^{\mathrm{A}},{\mv{H}}_{j+1}^{\mathrm{A}},\!\cdots\!,{\mv{H}}_{J}^{\mathrm{A}}]\!\in\! \mathbb{C}^{M\times \sum_{i\neq j}|\Lambda_i|}$ as the aggregate channel matrix from the UAV to all the available GBSs that receive a different data stream from $j$. Similarly, denote ${\mv{H}}_{[-j]}^{\mathrm{O}}\in \mathbb{C}^{M\times |\Psi_j|}$ as the aggregate channel matrix from the UAV to the set of occupied GBSs in $\Psi_j$. It then follows that the beamforming vector ${\mv{w}}_j$ should lie in the null space of ${\mv{H}}_{[-j]}=\left[{\mv{H}}_{[-j]}^{\mathrm{A}},{\mv{H}}_{[-j]}^{\mathrm{O}}\right]$, namely,
\begin{equation}\label{wj}
{\mv{w}}_j^H{\mv{H}}_{[-j]}={\mv{0}},\quad \forall j.
\end{equation}

\subsection{Comparison with Other Techniques}

Note that this paper utilizes the one-hop backhaul connections between adjacent GBSs to achieve partial interference cancellation. Alternatively, we can achieve the full interference mitigation via CoMP by connecting each occupied GBS to all the available GBSs via the backhaul links, i.e., $\Phi_{n_1}\!=\!\mathcal{N}_2,\forall n_1$, or connecting all the GBSs to a central decoding unit via fronthaul links as in C-RAN. Alternatively, we can achieve local interference cancellation via the uplink NOMA without any need for the backhaul/fronthaul signal forwarding, i.e., $\Phi_{n_1}\!=\!\emptyset,\forall n_1$. Specifically, with CoMP, as long as a data stream is decoded, it can be cancelled for decoding all the other data streams. As a result, we have the following proposition.
\begin{proposition}\label{proposition1}
With CoMP in which each occupied GBS has backhaul connections to all the available GBSs, i.e., $\Phi_{n_1}=\mathcal{N}_2$, $\forall n_1$, the maximum DoF of the multi-beam UAV uplink communication is $\min(M,N_2)$.
\end{proposition}
\begin{IEEEproof}
With $\Phi_{n_1}=\mathcal{N}_2$, we have $\Psi_j=\emptyset$ and thus $|\Psi_j|=0$, $\forall j$ according to (\ref{eqn:occupied GBS without data stream j}). It then follows from Theorem \ref{theorem1} that $J=\min(M,N_2)$ is achievable.
\end{IEEEproof}

It is worth noting that the above DoF is achieved at the expense of connecting all GBSs together via a large number of backhaul links, which is not cost-effective in practice and also introduces long delay which may not be amenable to the dynamic coverage area of the UAV due to its high mobility. In this regard, uplink NOMA may be appealing as no backhaul connections are required to perform local interference cancellation. Alternatively, the simple cognitive beamforming can be applied, which does not require any interference cancellation. For the purpose of DoF analysis, in this case we should adopt ZF-based transmit beamforming such that zero-interference is achieved at all occupied GBSs.  The DoF of the uplink NOMA scheme and the cognitive beamforming without cooperative interference cancellation is characterized by the following proposition.
\begin{proposition}\label{proposition2}
If all the GBSs are isolated without the exchange of decoded messages via backhaul links, i.e., $\Phi_{n_1}=\emptyset$, $\forall n_1$, the maximum DoF of the considered system is $\min(\max(M-N_1,0),N_2)$.
\end{proposition}
\begin{IEEEproof}
With $\Phi_{n_1}=\emptyset$, we have $\Psi_j=\mathcal{N}_1$ and thus $|\Psi_j|=N_1$, $\forall j$ according to (\ref{eqn:occupied GBS without data stream j}). It then follows from Theorem \ref{theorem1} that the maximum achievable DoF is $\max(\min(M-N_1,N_2),0)=\min(\max(M-N_1,0),N_2)$, for the case of ZF-based cognitive beamforming. This result also holds for uplink NOMA as successive interference cancellation (SIC) only improves the SINR, but does not help increase the DoF.\footnote{Note that there is an implicit assumption made here for the DoF analysis of uplink NOMA: the terrestrial users served by the occupied GBSs also have asymptotically high SNR as the UAV. This is for fair comparison with the other schemes considered in this section, where a stringent zero-interference condition holds at all the occupied GBSs, even when their associated terrestrial users have infinite SNR.}
\end{IEEEproof}

To summarize, our scheme based on cooperative interference cancellation can achieve a DoF gain over NOMA/cognitive beamforming with a reasonable requirement of the backhaul connections. Furthermore, from a practical system implementation viewpoint, the message-forwarding assisted interference cancellation in our proposed scheme is also more robust to the error propagation issue in NOMA due to the SIC.

\section{Proposed Solution at Finite SNR}\label{sec:Proposed Solution in Finite SNR Regime}
In this section, we focus on solving problem (\ref{eqn:P1}) in the finite SNR regime. As discussed in Section \ref{sec:UAV Transmission Rate Maximization with Interference Control}, the main difficulty lies in the coupled design of $J$, $\Lambda_j$'s, $\mv{w}_j$'s, and $R_j$'s, in which $J$ and $\Lambda_j$'s are discrete. To reduce the complexity for solving problem (\ref{eqn:P1}), in this section we propose to decouple the design of $J$ and $\Lambda_j$'s with that of $\mv{w}_j$'s and $R_j$'s.

\subsection{Data Stream Association Design}
In the first step, we aim to optimize $J$ and $\Lambda_j$'s. The key issue is that what criterion should be used to design $J$ and $\Lambda_j$'s. In Section \ref{sec:Proposed Solution}, we have shown in Theorem \ref{theorem1} that the maximum DoF of our considered system can be characterized by $\Lambda_j$'s solely. This paper thus selects $J$ and $\Lambda_j$'s that maximize the DoF of our considered system as the solution to problem (\ref{eqn:P1}). Such a suboptimal solution of $J$ and $\Lambda_j$'s decouples from the design of $\mv{w}_j$'s and $R_j$'s; thus, it not only reduces the algorithm complexity, but also achieves a reasonable performance since the DoF maximization problem (based on which we determine $J$ and $\Lambda_j$'s) is a good approximation to problem (\ref{eqn:P1}) at high SNR, which is usually the case for our considered multi-stream UAV uplink transmission for applications with high data rate (e.g., HD video) over strong LoS channels.

\subsection{Beamforming and Rate Allocation Design}
Next, given this solution of $J$ and $\Lambda_j$'s, problem (\ref{eqn:P1}) reduces to the following problem
\begin{subequations}\label{eqn:P4}\begin{align}
\mathop{ \underset{\{\mv{w}_j,R_j\}}{\max}} & ~ \sum\limits_{j=1}^J  R_j \label{eqn:problem 4}\\
\mathrm{s.t.} \ \ \ & ({\rm \ref{eqn:SINR constraint}}), ~ ({\rm \ref{eqn:interference temperature constraint}}), ~ ({\rm \ref{eqn:power constraint}}),
\end{align}\end{subequations}based on which we can jointly optimize $\mv{w}_j$'s and $R_j$'s to maximize the UAV sum-rate subject to the interference temperature constraints at the occupied GBSs.

It is worth noting that for information broadcasting applications in which a multi-antenna transmitter sends an independent message to each of the receivers simultaneously, the sum-rate maximization problem has been well-studied in the literature (see, e.g., \cite{WMMSE,optimality}). However, the results for sum-rate maximization in the MIMO broadcast channel (BC) cannot be used for solving problem (\ref{eqn:P4}) due to the minimum-rate operation in (\ref{eqn:rate}). In fact, our considered system can be considered as a multi-group multicast channel \cite{multicasting}, where each of the $J$ data streams is deemed to be decoded by an exclusive subset of the available GBSs. To the best knowledge of the authors, for a multi-group multicast channel, our considered sum-rate maximization problem is still open in the literature even without the interference temperature constraint (\ref{eqn:interference temperature constraint}), although a different transmit power minimization problem subject to the individual SINR constraints of each user has been investigated in \cite{multicasting}.

The objective function of problem (\ref{eqn:P4}) is a linear function over $R_j$'s. Moreover, the interference temperature constraint (\ref{eqn:interference temperature constraint}) and transmit power constraint (\ref{eqn:power constraint}) are also convex. However, the rate constraint (\ref{eqn:SINR constraint}) is non-convex. In the following, we apply the successive convex approximation technique to deal with the non-convex rate constraint (\ref{eqn:SINR constraint}) in problem (\ref{eqn:P4}).

By introducing a set of auxiliary variables $\eta_{n_2}$'s, it can be shown that the rate constraint (\ref{eqn:SINR constraint}) is equivalent to the following constraints:
\begin{align}
& |\mv{h}_{n_2}^H\mv{w}_j|^2\geq (2^{R_j}-1)\eta_{n_2}, ~~~ \forall n_2\in \Lambda_j, \forall j, \label{eqn:SINR 1} \\
& \sum\limits_{i\neq j} |\mv{h}_{n_2}^H\mv{w}_i|^2+\sigma_{n_2}^2\leq \eta_{n_2}, ~~~ \forall n_2\in \Lambda_j, \forall j. \label{eqn:SINR 2}
\end{align}Here, $\eta_{n_2}$ can be interpreted as the interference temperature constraint at the available GBS $n_2$ for decoding $s_j$. Constraint (\ref{eqn:SINR 2}) is a convex constraint. In the following, we deal with the non-convex constraint (\ref{eqn:SINR 1}).

First, it can be shown that by introducing some auxiliary variables $a_{n_2,j}$'s and $b_{n_2,j}$'s, constraint (\ref{eqn:SINR 1}) is equivalent to the following constraints:
\begin{align}
& a_{n_2,j}^2+b_{n_2,j}^2- (2^{R_j}-1)\eta_{n_2}\geq 0, ~~~ \forall n_2\in \Lambda_j, \forall j, \label{eqn:SINR 3} \\
& a_{n_2,j}={\rm Re}(\mv{h}_{n_2}^H\mv{w}_j), ~ b_{n_2,j}={\rm Im}(\mv{h}_{n_2}^H\mv{w}_j), ~~~ \forall n_2\in \Lambda_j, \forall j. \label{eqn:SINR 4}
\end{align}Next, given any $\tilde{a}_{n_2,j}$, $\tilde{b}_{n_2,j}$, and $\tilde{c}_{n_2,j}>0$, define a function of $a_{n_2,j}$, $b_{n_2,j}$, $R_j$, and $\eta_{n_2}$ as
\begin{align}
& f(a_{n_2,j},b_{n_2,j},R_j,\eta_{n_2}|\tilde{a}_{n_2,j},\tilde{b}_{n_2,j},\tilde{c}_{n_2,j}) \nonumber \\=&2\tilde{a}_{n_2,j}a_{n_2,j}+2\tilde{b}_{n_2,j}b_{n_2,j} -\tilde{a}_{n_2,j}^2-\tilde{b}_{n_2,j}^2-\left(\frac{\eta_{n_2}\tilde{c}_{n_2,j}}{2}+\frac{2^{R_j}-1}{2\tilde{c}_{n_2,j}}\right)^2. \label{eqn:function}
\end{align}It can be shown that $f(a_{n_2,j},b_{n_2,j},R_j,\eta_{n_2}|\tilde{a}_{n_2,j},\tilde{b}_{n_2,j},\tilde{c}_{n_2,j})$ is a concave function over $a_{n_2,j}$, $b_{n_2,j}$, $R_j$, and $\eta_{n_2}$. Moreover, the following inequality holds for $f(a_{n_2,j},b_{n_2,j},R_j,\eta_{n_2}|\tilde{a}_{n_2,j},\tilde{b}_{n_2,j},\tilde{c}_{n_2,j})$:
\begin{align}\label{eqn:inequality 1}
a_{n_2,j}^2+b_{n_2,j}^2-(2^{R_j}-1)\eta_{n_2}\geq f(a_{n_2,j},b_{n_2,j},R_j,\eta_{n_2}|\tilde{a}_{n_2,j},\tilde{b}_{n_2,j},\tilde{c}_{n_2,j}), ~~~ \forall n_2\in \Lambda_j, \forall j,
\end{align}where the equality holds if and only if $a_{n_2,j}=\tilde{a}_{n_2,j}$, $b_{n_2,j}=\tilde{b}_{n_2,j}$, and $\tilde{c}_{n_2,j}=\sqrt{(2^{R_j}-1)/\eta_{n_2}}$. Thus, we can use this concave lower bound $f(a_{n_2,j},b_{n_2,j},R_j,\eta_{n_2}|\tilde{a}_{n_2,j},\tilde{b}_{n_2,j},\tilde{c}_{n_2,j})$ to approximate constraint (\ref{eqn:SINR 3}) as follows:
\begin{align}\label{eqn:SINR 3 approximate}
f(a_{n_2,j},b_{n_2,j},R_j,\eta_{n_2}|\tilde{a}_{n_2,j},\tilde{b}_{n_2,j},\tilde{c}_{n_2,j})\geq 0, ~~~ \forall n_2\in \Lambda_j, \forall j.
\end{align}

Given any $\tilde{a}_{n_2,j}$'s, $\tilde{b}_{n_2,j}$'s, and $\tilde{c}_{n_2,j}>0$'s, with constraint (\ref{eqn:SINR constraint}) replaced by (\ref{eqn:SINR 2}), (\ref{eqn:SINR 4}) and (\ref{eqn:SINR 3 approximate}), we can solve the following approximated convex problem:
\begin{subequations}\label{eqn:P5}\begin{align}
\mathop{\underset{\{\mv{w}_j,R_j\},\{a_{n_2,j},b_{n_2,j}\},\{\eta_{n_2}\}}{\max}} & ~ \sum\limits_{j=1}^J R_j  \\
\mathrm{s.t.} \ \ \ \ \ \ \ \ \ \ & ({\rm \ref{eqn:interference temperature constraint}}), ~ ({\rm \ref{eqn:power constraint}}), ~ (\ref{eqn:SINR 2}), ~ (\ref{eqn:SINR 4}), ~ (\ref{eqn:SINR 3 approximate}).
\end{align}\end{subequations}
Problem (\ref{eqn:P5}) is a convex optimization problem, thus can be solved efficiently by CVX \cite{CVX}.

After solving problem (\ref{eqn:P5}) given any point $\tilde{a}_{n_2,j}$'s, $\tilde{b}_{n_2,j}$'s, and $\tilde{c}_{n_2,j}>0$'s, the successive convex approximation method for solving problem (\ref{eqn:P4}) proceeds by iteratively updating $\tilde{a}_{n_2,j}$'s, $\tilde{b}_{n_2,j}$'s, and $\tilde{c}_{n_2,j}>0$'s based on the solution to problem (\ref{eqn:P5}). The proposed iterative algorithm is summarized in Algorithm \ref{table1}, where $q$ denotes the index of iteration, and $\epsilon>0$ is a small value to control the convergence of the algorithm. The convergence of Algorithm \ref{table1} is guaranteed by the following theorem.

\begin{algorithm}[t]
{\bf Initialization}: Set the initial values for $\tilde{a}_{n_2,j}$'s, $\tilde{b}_{n_2,j}$'s, and $\tilde{c}_{n_2,j}>0$'s and $q=1$; \\
{\bf Repeat}:
\begin{enumerate}
\item Find the optimal solution to problem (\ref{eqn:P5}) using CVX as $\{\mv{w}_j^{(q)},R_j^{(q)},a_{n_2,j}^{(q)},b_{n_2,j}^{(q)},\eta_{n_2}^{(q)}\}$;
\item Update $a_{n_2,j}=\tilde{a}_{n_2,j}^{(q)}$, $b_{n_2,j}=\tilde{b}_{n_2,j}^{(q)}$, and $\tilde{c}_{n_2,j}=\sqrt{(2^{R_j^{(q)}}-1)/\eta_{n_2}^{(q)}}$;
\item $q=q+1$.
\end{enumerate}
{\bf Until} $\sum_{j=1}^J R_j^{(q)}-\sum_{j=1}^J R_j^{(q-1)}<\epsilon$.
\caption{Proposed Algorithm for Solving Problem (\ref{eqn:P4}).}
\label{table1}
\end{algorithm}

\begin{theorem}\label{theorem2}
Monotonic convergence of Algorithm \ref{table1} is guaranteed, i.e., $\sum_{j=1}^J R_j^{(q)}\geq \sum_{j=1}^J R_j^{(q-1)}$, $\forall q\geq 2$. Moreover, the converged solution satisfies all the Karush-Kuhn-Tucker (KKT) conditions of problem (\ref{eqn:P4}).
\end{theorem}
\begin{IEEEproof}
The proof of Theorem \ref{theorem2} directly follows that for \cite[Theorem 1]{SCA}, and is thus omitted for brevity.
\end{IEEEproof}

The overall algorithm to solve problem (\ref{eqn:P1}) with a joint optimization of $J$, $\Lambda_j$'s, $\mv{w}_j$'s, and $R_j$'s is summarized in Algorithm \ref{table2}.

\begin{algorithm}[t]
\begin{enumerate}
\item Find the maximum DoF $J$ such that there exists a feasible solution of the data assignment $\Lambda_j$'s to problem (\ref{eqn:P3}), and fix the obtained $J$ and $\Lambda_j$'s as the solution to problem (\ref{eqn:P1});
\item Find the beamforming solution $\mv{w}_j$'s and rate allocation solution $R_j$'s to problem (\ref{eqn:P1}) via Algorithm \ref{table1}.
\end{enumerate}
\caption{Proposed Algorithm for Solving Problem (\ref{eqn:P1}).}
\label{table2}
\end{algorithm}

\section{Numerical Examples}\label{sec:Numerical Examples}
In this section, we provide numerical examples to verify the performance of the proposed cooperative interference cancellation strategy for the multi-beam UAV communication in the cellular uplink. We consider the cellular network topology shown in Fig. \ref{cell}, where there are $N=8$ GBSs, among which $N_1=3$ GBSs are occupied GBSs with $\mathcal{N}_1=\{1,2,3\}$, and $N_2=5$ GBSs are available GBSs with $\mathcal{N}_2=\{4,5,6,7,8\}$. Each GBS covers a cell with a radius of $200$ meters (m). Moreover, the backhaul connections between the occupied GBSs and the available GBSs are specified by $\Phi_1=\{4,5,6\}$, $\Phi_2=\{5,6,7\}$, and $\Phi_3=\{6,7,8\}$. The UAV is assumed to be right on the top of the center of cell 6, and its altitude is $100$ m. We assume that in the Rician fading channel model (\ref{eqn:Rician channel}), the Rician factor is $\lambda=5$, and the LoS component $\hat{\mv{h}}_n$ follows the linear antenna array model \cite{Luo07}. At last, the bandwidth of the RB used by the UAV is $10$ MHz, while the power spectrum density of the AWGN at the GBSs is $-169$ dBm/Hz.

\subsection{Achievable DoF}
First, we check the DoF of this setting achieved by our proposed cooperative interference cancellation strategy as characterized in Theorem \ref{theorem1}. The maximum achievable DoF for our proposed scheme is obtained by solving problem (\ref{eqn:P2}). In the following, we list the optimal solution to problem (\ref{eqn:P2}) for various values of $M$.
\begin{enumerate}
\item $M=1$, $M=2$: Maximum DoF $J=1$ is achievable with $\Lambda_1=\{6\}$.
\item $M=3$: Maximum DoF $J=2$ is achievable with $\Lambda_1=\{4,6\}$, $\Lambda_2=\{5,7\}$.
\item $M=4$, $M=5$: Maximum DoF $J=3$ is achievable with $\Lambda_1=\{5\}$, $\Lambda_2=\{6\}$, and $\Lambda_3=\{7\}$.
\item $M=6$: Maximum DoF $J=4$ is achievable with $\Lambda_1=\{4,8\}$, $\Lambda_2=\{5\}$, $\Lambda_3=\{6\}$, and $\Lambda_4=\{7\}$.
\item $M\geq7$: Maximum DoF $J=\min(M,N_2)=5$ is achievable with $\Lambda_1=\{4\}$, $\Lambda_2=\{5\}$, $\Lambda_3=\{6\}$, $\Lambda_4=\{7\}$, and $\Lambda_5=\{8\}$.
\end{enumerate}
Note that the optimal solution to problem (P1) may not be unique. For example, when $M=1$, $\Lambda_1=\{5,7\}$ is also an optimal solution.
\begin{figure}[t]
  \centering
  \includegraphics[width=8cm]{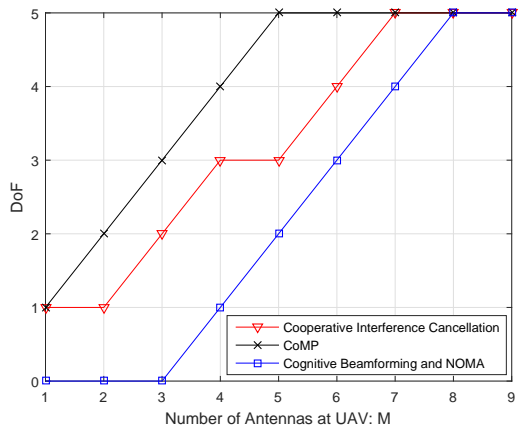}
  \caption{Maximum achievable DoF versus number of antennas at the UAV.}\label{DoF}
\end{figure}

For comparison, we also evaluate the maximum DoF achieved by the CoMP and ZF-based cognitive beamforming (or uplink NOMA), as characterized in Propositions \ref{proposition1} and \ref{proposition2}, respectively. In Fig. \ref{DoF}, we show the maximum achievable DoF versus the number of antennas at the UAV, i.e., $M$, achieved by our proposed scheme and the benchmark schemes. It can be observed that as compared to the performance upper bound, i.e., CoMP with full cooperation, our proposed partial cooperative interference cancellation scheme achieves a reasonable DoF with various values of $M$. However, our scheme can be easily implemented in practice thanks to the recent development of Xhaul between adjacent GBSs for information exchange \cite{Xhaul}, while an ideal CoMP architecture is still difficult to implement in practice. On the other hand, as compared to the cases of ZF-based cognitive beamforming without information exchange between adjacent GBSs and uplink NOMA with local SIC at the occupied GBSs, it is observed that our proposed scheme achieves significant DoF gain, especially when the number of antennas at the UAV is small.

\subsection{Sum-Rate Maximization at Finite SNR}
Next, we consider the UAV's achievable sum-rate in the finite SNR regime. Assume that the UAV has $M=5$ antennas. As shown in Fig. \ref{DoF}, the maximum DoF is $J=3$ when $M=5$. However, there are several data stream association solutions to achieve the maximum DoF. In the following, we consider two solutions: all the available GBSs $4-8$ are utilized, i.e., $\Lambda_1=\{4,7\}$, $\Lambda_2=\{5,8\}$, $\Lambda_3=\{6\}$; and only the available GBSs $5-7$ are utilized, i.e., $\Lambda_1=\{5\}$, $\Lambda_2=\{6\}$, $\Lambda_3=\{7\}$. It can be observed that under the first solution, we have $\Gamma_{n_1}=\emptyset$, $\forall n_1\in \mathcal{N}_1$, in problem (\ref{eqn:P4}), and under the second solution, we have $\Gamma_1=\{3\}$, $\Gamma_2=\emptyset$, $\Gamma_3=\{1\}$ in problem (\ref{eqn:P4}).

\subsubsection{Convergence of Algorithm \ref{table1}}

\ \ \

\begin{figure}[t]
  \centering
  \includegraphics[width=8cm]{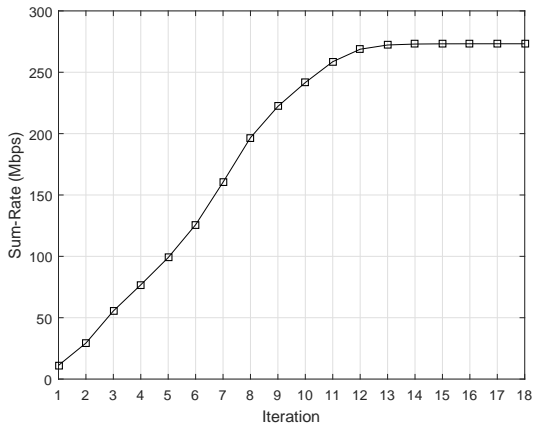}
  \caption{Convergence of Algorithm \ref{table1}.}\label{convergence}
\end{figure}

First, we verify the convergence of Algorithm \ref{table1}. The transmit power constraint of UAV is $23$ dBm. Moreover, we merely consider the data stream association solution $\Lambda_1=\{4,7\}$, $\Lambda_2=\{5,8\}$, $\Lambda_3=\{6\}$. Fig. \ref{convergence} shows the sum-rate versus the number of iterations in Algorithm \ref{table1}. A monotonic convergence is observed from Fig. \ref{convergence}, which verifies Theorem \ref{theorem2}.

\subsubsection{Performance Comparison}

\ \ \

Next, we briefly introduce some benchmark schemes for UAV transmission rate maximization with interference control to the occupied GBSs, and then demonstrate the effectiveness of the cooperative interference cancellation strategy in Section \ref{sec:Cooperative Interference Cancellation} via performance comparison.

\textbf{Benchmark Scheme 1: CoMP.} Under this benchmark scheme, we consider the case when all the GBSs are fully connected with each other with backhaul links. Mathematically, such a scheme results in $\Gamma_{n_1}=\emptyset$, $\forall n_1\in \mathcal{N}_1$, in problem (\ref{eqn:P4}). In other words, the system reduces to a point-to-point MIMO channel with $M$ transmit antennas and $N_2$ receive antennas, the capacity of which can be achieved by channel singular value decomposition (SVD) based linear precoding and decoding together with water-filling based power control \cite{Cover}.

\textbf{Benchmark Scheme 2: Cognitive Beamforming \cite{Interference_Temperature}.} Under this benchmark scheme, we consider the cognitive beamforming scheme \cite{Interference_Temperature} to control interference leakage via beamforming design alone, without utilization of the backhaul links for cooperative interference cancellation.  Mathematically, such a scheme results in $\Gamma_{n_1}=\{1,\cdots,J\}$, $\forall n_1\in \mathcal{N}_1$, in problem (\ref{eqn:P4}). Since the DoF in this setting is $J=M-N_1=2$, we assume that two data streams are sent. Then, Algorithm \ref{table1} can be used to design the beamforming vectors by solving problem (\ref{eqn:P4}) with $\Gamma_{n_1}=\{1,\cdots,J\}$, $\forall n_1\in \mathcal{N}_1$.

\begin{figure}[t]
  \centering
  \includegraphics[width=8cm]{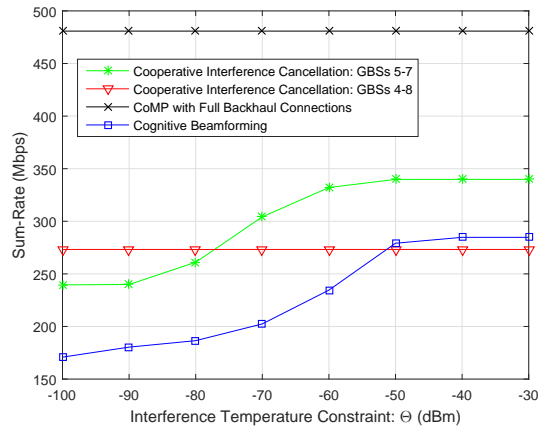}
  \caption{Sum-rate versus interference temperature constraint when the transmit power constraint is $23$ dBm.}\label{fig_rate}
\end{figure}

Fig. \ref{fig_rate} shows the sum-rate achieved by the proposed strategy as well as the benchmark schemes, with various interference temperature constraints, where the UAV transmit power constraint is $23$ dBm. It is observed that under the proposed strategy, utilizing all the available GBSs for cooperative interference cancellation achieves higher sum-rate when the interference temperature constraint is stringent, i.e., below $-80$ dBm, but achieves lower sum-rate when the interference temperature constraint is not stringent, i.e., above $-70$ dBm. This is because if the data stream association solution $\Lambda_1=\{4,7\}$, $\Lambda_2=\{5,8\}$, $\Lambda_3=\{6\}$ is adopted, the interference temperature constraint can be eliminated when we design the UAV beamforming vectors, but data streams $1$ and $2$ need to be multicast to two GBSs (thus, leading to less beamforming gain); while if the data stream association solution $\Lambda_1=\{5\}$, $\Lambda_2=\{6\}$, $\Lambda_3=\{7\}$ is adopted, each data stream is only sent to one close GBS, but the interference temperature constraint needs to be considered when we design the UAV beamforming vectors since the occupied GBSs $1$ and $3$ cannot receive all the data streams for interference cancellation via the backhaul links (thus, resulting in less interference cancellation gain). As a result, the data stream association should be carefully designed in practice to balance the beamforming gain and interference cancellation gain. Next, it is observed that CoMP achieves the maximum sum-rate thanks to the fully cooperative interference cancellation. However, this scheme requires a massive number of backhaul links to connect all th GBSs, which is difficult to realize in practice, while our strategy merely requires the backhaul connections between adjacent cells, which can be realized by the current Xhaul structure \cite{Xhaul}. At last, it is observed that the sum-rate achieved by our proposed strategy is much higher than that achieved by the cognitive beamforming thanks to that of the cooperative interference cancellation gain.

\begin{figure}[t]
  \centering
  \includegraphics[width=8cm]{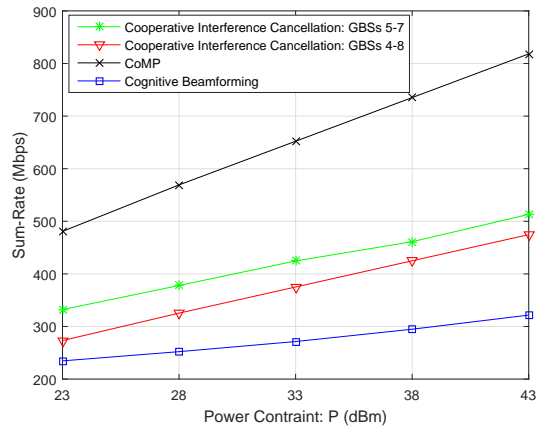}
  \caption{Sum-rate versus transmit power constraint when the interference temperature constraint is $-60$ dBm.}\label{fig_rate_power}
\end{figure}

Fig. \ref{fig_rate_power} shows the sum-rate achieved by the proposed strategy as well as the benchmark schemes, with various transmit power constraints, where the interference temperature constraint is $-60$ dBm. It is observed that at the high SNR regime, the sum-rate performance is consistent with the DoF performance. Specifically, the DoF achieved by our proposed scheme and Benchmark Schemes 1 and 2 is $3$, $5$, and $2$, respectively. As a result, the sum-rate achieved by CoMP and our proposed scheme increases faster with the transmit power as compared to the cognitive beamforming scheme.

\section{Conclusion}\label{sec:Conclusion}
This paper studied the uplink communication from a multi-antenna UAV to multiple GBSs. To achieve high data rate while effectively controlling the severe aerial interference to the uplink terrestrial communications arising from the strong LoS channel between the UAV and GBSs, this paper proposed a multi-beam UAV communication scheme with a novel cooperative interference cancellation strategy, which exploits the use of existing backhaul links among adjacent GBSs in the cellular network. Specifically, the UAV sends multiple data streams to a subset of available GBSs that are not using the same RB for terrestrial communications, which then forward the decoded messages to the backhaul-connected occupied GBSs for canceling the UAV's interference before decoding the messages of their associated terrestrial users. The maximum achievable DoF of the proposed strategy was characterized by optimizing the association between each data stream and its decoding available GBSs. Based on this data stream association, the sum-rate of the UAV uplink communication was maximized at finite SNR by a proper beamforming design. Numerical examples showed that significant DoF and sum-rate gains can be achieved with our proposed strategy compared to benchmark schemes with no or local interference cancellation. In future work, we will extend the DoF and/or sum-rate optimization for our proposed scheme to the cases of multi-antenna GBSs with flexible 3D beamforming and uplink NOMA with both cooperative and local interference cancellation.

\begin{appendix}
\subsection{Proof of Theorem \ref{theorem1}}\label{appendix1}
First, we show that if a solution of $\Lambda_j$'s satisfies conditions (\ref{eqn:message constraint}), (\ref{eqn:one data stream}), and (\ref{eqn:null space}), then we can find $J$ beamforming vectors, $\mv{w}_j$'s, $j=1,\cdots,J$, that satisfy conditions (\ref{eqn:DoF 1}), (\ref{eqn:DoF 3}), (\ref{eqn:DoF 2}), (\ref{eqn:message constraint}), and (\ref{eqn:one data stream}). For convenience, define $\mv{H}_{{\rm occ},j}=[\cdots,\mv{h}_{n_1},\cdots,\forall n_1\in \Psi_j]$ and $\mv{H}_{{\rm ava},j}=[\cdots,\mv{h}_{n_2},\cdots,\forall n_2\in \bigcup_{i\neq j}\Lambda_i]$, $j=1,\cdots,J$. According to (\ref{eqn:DoF 3}) and (\ref{eqn:DoF 2}), each beamforming vector $\mv{w}_j$ should lie in the null space of $[\mv{H}_{{\rm occ},j},\mv{H}_{{\rm ava},j}]$. If (\ref{eqn:null space}) holds, then for each $1\leq j \leq J$, we have ${\rm rank}([\mv{H}_{{\rm occ},j},\mv{H}_{{\rm ava},j}])\leq {\rm rank}(\mv{H}_{{\rm occ},j})+{\rm rank}(\mv{H}_{{\rm ava},j})\leq |\Psi_j|+\sum\limits_{i\neq j}|\Lambda_i|<M$. In other words, the null space of $[\mv{H}_{{\rm occ},j},\mv{H}_{{\rm ava},j}]$ exists for each $j$, i.e., we can find $J$ beamforming vectors to satisfy (\ref{eqn:DoF 3}) and (\ref{eqn:DoF 2}). Moreover, under the Rician channel model as shown in (\ref{eqn:Rician channel}), $\mv{h}_n$'s are linearly independent with each other. Since the ZF beamforming vectors only depend on $\mv{H}_{{\rm occ},j}$'s and $\mv{H}_{{\rm ava},j}$'s but are not related to $\mv{h}_{n_2}$'s, $n_2\in \Lambda_j$, with probability one these ZF beamforming vectors do not lie in the null space of $\mv{h}_{n_2}$'s, $n_2\in \Lambda_j$, i.e., (\ref{eqn:DoF 1}) holds.

Next, we show that if we can find $J$ beamforming vectors, $\mv{w}_j$'s, $j=1,\cdots,J$, that satisfy conditions (\ref{eqn:DoF 1}), (\ref{eqn:DoF 3}), (\ref{eqn:DoF 2}), (\ref{eqn:message constraint}), and (\ref{eqn:one data stream}), then conditions (\ref{eqn:message constraint}), (\ref{eqn:one data stream}), and (\ref{eqn:null space}) must hold. Note that with probability one we have ${\rm rank}([\mv{H}_{{\rm occ},j},\mv{H}_{{\rm ava},j}])=M$ if $|\Psi_j|+\sum\limits_{i\neq j}|\Lambda_i|\geq M$ for some $j$, since $\mv{h}_n$'s are linearly independent with each other. In this case, the null space of $[\mv{H}_{{\rm occ},j},\mv{H}_{{\rm ava},j}]$ does not exist for this particular $j$, and it is impossible to design the ZF beamforming vector.

Theorem \ref{theorem1} is thus proved.

\end{appendix}

\bibliographystyle{IEEEtran}
\bibliography{CIC}
\end{document}